\documentclass[manuscript]{acmart}
\usepackage{algpseudocode}
\usepackage{multirow}
\usepackage[linesnumbered,ruled,vlined]{algorithm2e}

\begin{document}
\title{Prompt Tuning Large Language Models on Personalized Aspect Extraction for Recommendations} % To be decided 
% \title{Personalized Aspect Extraction for Explainable Recommendations} % To be decided 
% \title{Continuous Prompt Tuning for Personalized Aspect-Based Recommendations}
%\title{Continuous Prompt Tuning for Providing Aspect-Based Recommendations}

%\author{Anonymous for Review}

\author{Pan Li}
\affiliation{%
  \institution{New York University}
  \streetaddress{44 West 4th Street}
  \city{New York}
  \country{USA}}
\email{pli2@stern.nyu.edu}

\author{Yuyan Wang}
\affiliation{%
  \institution{Google Research}
  \streetaddress{Mountain View}
  \city{California}
  \country{USA}}
\email{yuyanw@google.com}

\author{Ed H. Chi}
\affiliation{%
  \institution{Google Research}
  \streetaddress{Mountain View}
  \city{California}
  \country{USA}}
\email{edchi@google.com}

\author{Minmin Chen}
\affiliation{%
  \institution{Google Research}
  \streetaddress{Mountain View}
  \city{California}
  \country{USA}}
\email{minminc@google.com}

\begin{abstract}
There have been growing interests in providing explainable recommendations for a frictionless user experience on personalization platforms~\cite{zhang2020explainable}. We focus on the branch of recommendation explanation that is opinion-based or more specifically aspect-based, i.e,  understanding user preference and recommendation along multiple interpretable dimensions. Existing works on this topic mainly focus on the following two goals: aspect extraction and aspect-based recommendation. On one hand, existing aspect extraction methods mostly rely on explicit or ground truth aspect information, or using data mining or machine learning approaches to extract aspects from implicit user feedback such as user reviews. It however remains under-explored how the extracted aspects can help generate more meaningful recommendations to the users. On the other, existing research on aspect-based recommendations often relies on separate aspect extraction models or assumes the aspects are given, without accounting for the fact the optimal set of aspects could be dependent on the recommendation task at hand. 

In this work, we propose to combine aspect extraction together with aspect-based recommendations in an end-to-end manner, achieving the two goals together in a single framework. For the aspect extraction component, we leverage the recent advances in large language models and design a new prompt learning mechanism to generate aspects for the end recommendation task. For the aspect-based recommendation component, the extracted aspects are concatenated with the usual user and item features used by the recommendation model. The recommendation task mediates the learning of the user embeddings and item embeddings, which are used as soft prompts to generate aspects. Therefore, the extracted aspects are personalized and contextualized by the recommendation task. We showcase the effectiveness of our proposed method through extensive experiments on three industrial datasets, where our proposed framework significantly outperforms state-of-the-art baselines in both the personalized aspect extraction and aspect-based recommendation tasks. In particular, we demonstrate that it is necessary and beneficial to combine the learning of aspect extraction and aspect-based recommendation together. We also conduct extensive ablation studies to understand the contribution of each design component in our framework.

\end{abstract}

\keywords{Prompt Tuning, Aspect Learning, Aspect-based Recommendation, Recommender System}

\maketitle

\section{Introduction}
Recommender systems have been widely deployed in the industry to identify the most relevant content for the users \cite{adomavicius2005toward}. Explainable recommendations, which provide explanations along recommendations, have been shown to improve the transparency, persuasiveness, effectiveness, trustworthiness, and satisfaction of recommendation systems\cite{zhang2020explainable}. 
To this end, researchers have developed a series of explainable recommender systems, offering different forms of explanations such as textual, visual, or social explanations.

Among various strands of research on explainable recommendations, the aspect-based explanation method \cite{guan2019attentive} has attracted a lot of attention due to its simplicity and superior performance. "Aspect" refers to a specific dimension of user experience that may affect her/his decision-making process. For example, when a user is browsing the TripAdvisor website to select the hotel for an upcoming holiday trip, she/he might pay special attention to the location, price, service, or breakfast of the candidate hotels. These factors are all considered aspects of the hotel that should be accounted for when making recommendations.

Existing aspect-based recommendation methods \cite{bauman2017aspect} typically rely on available aspect terms in the datasets, and focus on identifying the relative importance of each aspect in the users' decision-making processes. However in practical applications, the aspect information is often not readily available, and we need to learn and extract such information from various types of user feedback, including ratings and reviews. On the other hand,  identifying aspects from text inputs has been actively studied in the NLP community \cite{xu2018double}.  Existing work however studies aspect extraction in isolation from downstream applications such as providing recommendations or understanding user preferences. We here argue that it is  beneficial to connect aspect-term extraction and aspect-based recommendation in an end-to-end manner. As a result, we will be able to generate aspect terms that not only are representative of the user reviews, but also are useful for making better recommendations for the users.

In particular, motivated by the recent success of  pre-trained large language models (LLMs) \cite{li2021prefix}, we explore using these models to perform aspect extraction. LLMs, pre-trained on the enormous text corpus available on the web, are powerful for different NLP tasks \cite{devlin2018bert,radford2019language,brown2020language}.
The benefits of using LLM for aspect extraction lie in three folds. First, LLM can utilize a large corpus from the pre-trained model to better understand the semantic information within user reviews, and therefore better identifies the key aspect terms. Second, as demonstrated in recent studies \cite{zhao2023survey,bubeck2023sparks}, LLM achieves significantly better performance in a wide range of NLP tasks, including Named Entity Recognition, over the state-of-the-art baseline models. Lastly, LLMs can perform NLP tasks with just a few examples demonstrating what needs to be done (through few-shot or even zero-shot learning) \cite{brown2020language}, while in practical recommendation applications, we only have very few ground-truth aspect information of user reviews.

Specifically, We propose an end-to-end framework where it combines the learning of LLM-based personalized aspect extraction through prompt tuning~\cite{lester-etal-2021-power} and aspect-based recommendation together to produce better recommendations.
To adapt the pre-trained LLMs, in particular GPT-2 for the aspect term extraction task, we fine-tune the LLM on our offline datasets and update the parameters of the LLM and the fine-tuning layer following the standard practice \cite{howard2018universal}. 
The soft prompts for the LLM are constructed based on user and item IDs/features, which are then concatenated with input embeddings from tokenizing and embedding the review text, and fed into the fine-tuned LLM. The output of the language model is a list of aspect terms, which are then fed into the attentive neural network along with user and item embeddings to generate aspect-based recommendations.

To showcase the efficacy of our proposed method, we conducted extensive offline experiments on three large-scale real-world  datasets, where our proposed model significantly outperforms selected state-of-the-art baseline methods in both aspect-term extraction and aspect-based recommendation tasks. We also study the effect of combining aspect learning with the aspect-based recommendation as proposed in our method. We find joint learning yields the generated aspects to be not only representative of user reviews and experiences, but also useful for producing better recommendations. We also conduct extensive ablation studies to understand the importance of each component in our proposed framework.

In summary, we make the following research contributions:
\begin{itemize}
\item We are the first to combine the aspect term extraction and aspect-based recommendation tasks in one single framework in an end-to-end learning manner. The proposed framework automatically extracts the most important personalized aspects from user reviews, and utilizes the extracted aspects to produce better recommendations.
\item We test our proposed method through extensive offline experiments on three public datasets, showcasing that our model can effectively capture the most important aspects in user reviews, and achieves significant performance improvements over state-of-the-art baselines.
\item We provide empirical evidence to demonstrate the importance of the joint training of aspect learning together with the aspect-based recommendation, in order to extract the most meaningful personalized aspects for better explainable recommendations.
\end{itemize}

\section{Related Work}
\subsection{Aspect Term Extraction}
Our work is related to the research in aspect term extraction in the field of natural language processing \cite{hu2004mining}, in both unsupervised and supervised fashion. The former group of methods include frequent pattern mining \cite{popescu2007extracting}, syntactic rules-based extraction \cite{zhuang2006movie,qiu2011opinion}, topic modeling \cite{mei2007topic,titov2008joint,moghaddam2011ilda}, word alignment \cite{liu2013opinion} and label propagation \cite{shu2016lifelong}. The latter group of methods learns to come up with the annotated aspect terms from the input using different model architectures, such as Conditional Random Fields (CRF) \cite{jakob2010extracting} and deep neural network techniques, such as LSTM \cite{liu2015fine}, CNN \cite{xu2018double}, and attention mechanism \cite{wang2017coupled,he2017unsupervised}. Furthermore, researchers in \cite{li2017deep} have also proposed aspect terms and opinion co-extraction via deep neural networks, leveraging ground truth aspect terms and sentiment lexicon for opinion extraction. 

While existing methods have been successful in extracting aspect terms in user reviews accurately and efficiently, they do not measure the effectiveness of the extracted aspects for downstream applications, for example, through aspect-based recommendations. As a result, while the extracted aspects could represent the essence of user reviews, they may or may not be beneficial for the subsequent recommendation process. In this work, we propose to combine the aspect term extraction and aspect-based recommendation tasks in an end-to-end manner and demonstrate that the joint training significantly improves the performance of both tasks.

\subsection{Aspect-Based Recommendation}
In parallel to aspect term extraction, recommendation researchers have also proposed a series of models to utilize the extracted aspect information to provide better recommendations. For example, \cite{ganu2013improving} manually defined a set of aspects and incorporated it into a regression-based method for rating prediction. \cite{chen2016learning} proposed a tensor-matrix factorization method to select the most interesting product aspects for each user with a learning-to-rank method. The rating scores were then predicted as the weighted summation of the product’s sentiment scores on the user’s most cared product aspects. \cite{bauman2017aspect} also extracted aspects with external tools and then train a latent factor model SLUM for every aspect to predict user sentiment scores toward each aspect of a product. In recent years, with the rapid development of deep learning techniques in NLP and data mining communities, there have also been a lot of neural network-based aspect recommendation models, such as A3NCF \cite{cheng20183ncf}, ALFM \cite{cheng2018aspect}, ANR \cite{chin2018anr} and AARM \cite{guan2019attentive}, which all aim at capturing users’ varied interests towards multi-dimensional aspects in user reviews.

Our proposed method differs from related works in multiple perspectives. To start with, existing works treat the aspect term extraction and aspect-based recommendations as two separate tasks, and rely heavily on existing or pre-trained aspect information in order to generate satisfying product recommendations. In practical industrial settings, such aspect information is usually difficult and costly to obtain. In addition, existing works on aspect extraction are not personalized or contextualized, and can as a result miss important aspects which are important in an individual user's decision-making process.  
To the best of our knowledge, there is no existing work that leverages the capability of pre-trained LLMs for understanding and extracting the multi-dimensional aspect information in user reviews. We propose a novel aspect-based recommendation method based on soft prompt tuning with LLMs, which could automatically extract the most important and useful aspects from user reviews in a personalized fashion, and utilize the extracted aspects to produce even better personalized recommendations.

\subsection{Pre-trained Language Model and Prompt Learning}
Recent years have witnessed the rise of pre-trained large language models, such as transformer \cite{vaswani2017attention} and BERT \cite{devlin2018bert}, which were first brought to the domain of natural language generation and machine translation with the encoder-decoder architecture. Pre-trained and fine-tuned large language models have shown to be effective on a wide range of natural language understanding tasks \cite{radford2018improving} such as commonsense reasoning and question answering \cite{wei2022chain}. We consider these models highly potential in understanding multi-dimensional user preferences and aspect information in textual reviews. Training large language models from scratch is extremely costly though. There has been a line of emerging research in prompt learning \cite{lester-etal-2021-power, liu2023pre} and fine-tuning \cite{liu2022p}, where researchers instead adapt pre-trained models to their tasks. Prompt learning has been successfully applied to many applications, such as domain adaptation \cite{ben2021pada}, text summarization \cite{li2021prefix}, and image captioning \cite{tsimpoukelli2021multimodal}, and has shown to be effective in certain sentiment analysis applications. In this paper, we adopt a soft prompt learning method to extract aspect terms from user reviews, and then connect the learning process with aspect-based recommender systems in an end-to-end manner.

\section{Method}
\begin{figure*}
\centering
\includegraphics[width=0.75\textwidth]{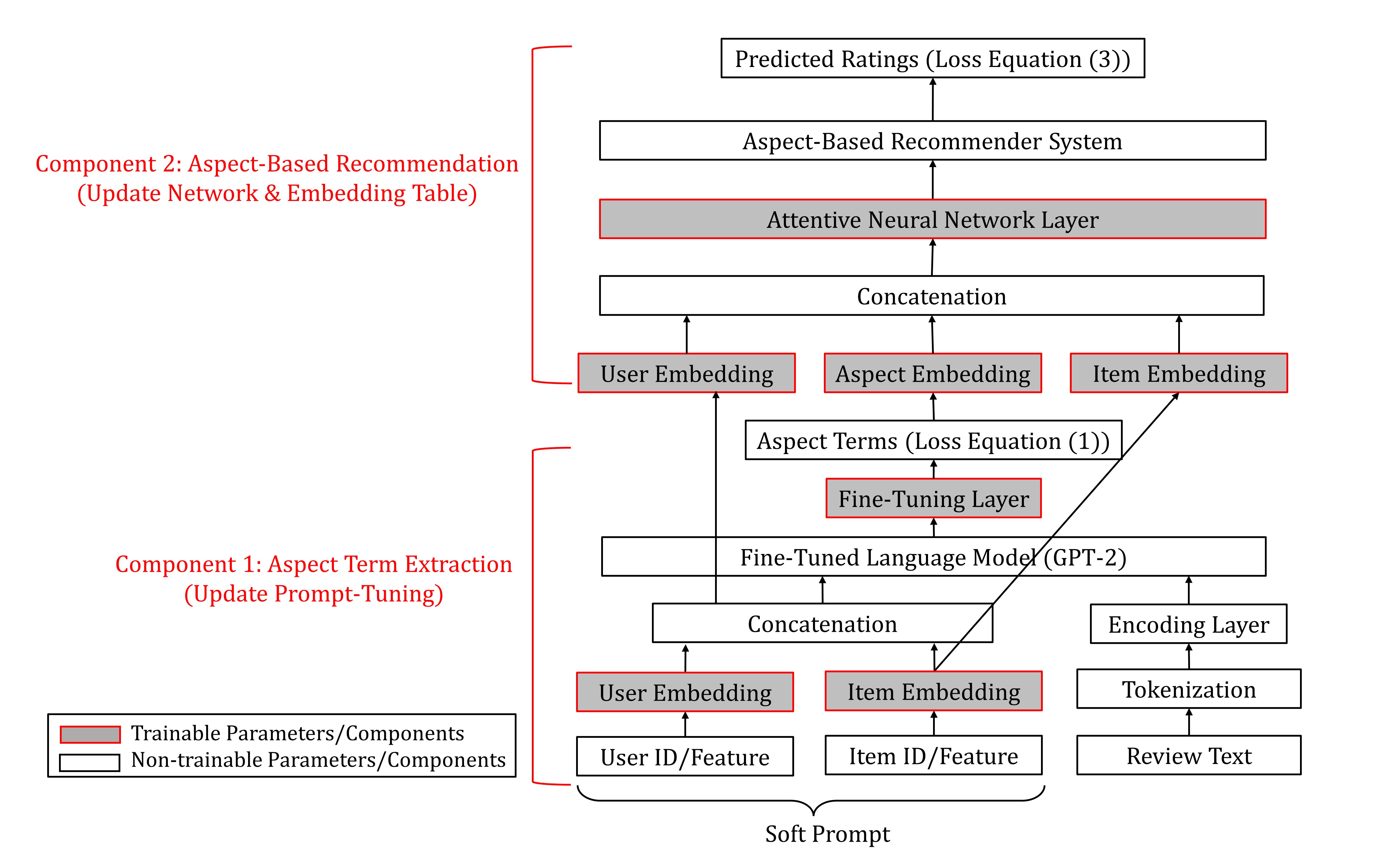}
\caption{Overview of the Proposed Model.}
\label{model}
\end{figure*}

In this section, we introduce the new aspect-based recommendation method based on soft prompt tuning, which could automatically extract the most important personalized aspects from user reviews, and utilize the extracted aspects to produce better recommendations. We first formulate the problem under the prompt tuning framework and then present the aspect-based recommendation workflow. An overview of the model is shown in Figure \ref{model}. As shown in the figure, the model has two components, where the first component is for aspect extraction, and the second is for aspect-based recommendation with the extracted aspects. In Section \ref{sec:aspect_extraction} and \ref{sec:aspect_based_recs}, we introduce the two components separately. The joint learning procedure of the two components is described in Section \ref{sec:joint_learning}.

\subsection{Component 1: Soft Prompt Tuning for Personalized Aspect Extraction}
\label{sec:aspect_extraction}
Pre-trained large language models, such as BERT \cite{devlin2018bert} and GPT-2 \cite{radford2019language}, trained on massive amounts of textual data and contain tens or hundreds of billions of model parameters, have revolutionized the field of natural language processing. LLMs achieved immense success and substantially out-perform previous state-of-the-art in many natural language processing tasks.
Due to the scale of these models, re-training them from scratch is extremely costly. Instead, people rely on prompt engineering (or priming) ~\cite{brown2020language} to modulate the behavior of these pre-trained LLMs for different tasks. Prompt tuning~\cite{lester-etal-2021-power, liu2023pre}, which learns soft prompts rather than relying on human-engineered prompts, has been shown to be much more efficient and robust in conditioning these pre-trained LLMs for the specific downstream tasks. We thus adopt the prompt tuning approach for aspect term extraction in our work.

We now describe the process of prompt learning for personalized aspect term extract as follows.
We first construct the prompt for instructing the language models, such as \emph{"This is the review written by user $u$ for item $i$. Please extract the top three aspect words in this review."}. This prompt is subsequently tokenized and embedded through a separate set of parameters, while the associated review $R$ is tokenized and embedded through the parameters of pre-trained LLMs. The embeddings are then concatenated together and fed into the (encoder-)decoder as normal.
The pre-trained LLM as a result is instructed to extract the top three aspect terms based on the given user $u$, item $i$, and review $R$. The performance of the prompt learning models will be determined by the following factors: (1) the quality of the prompt; (2) the quality of the input representations; and (3) the quality of the language model. We will now discuss the design choices of these three factors respectively to apply the prompt learning paradigm. 

\noindent\textbf{\emph{Constructing soft prompts.}} To manually design an optimal template that best fits the designated natural language processing tasks is usually the bottleneck for prompt learning-based methods. Fortunately, as shown in recent literature \cite{li2021prefix}, the prompt does not have to be sophisticated, and simple templates can achieve satisfying performance. In addition, it is shown in \cite{li2022personalized} that in many applications, we do not need to include the template words explicitly in the prompt, such as "Please extract the top three aspect words in this review.", which is identical for all the examples in the same task.
Instead, we only need to include the information that is useful for the personalized aspect learning process, e.g, user and item representations. Different users write reviews in different styles, and they also emphasize on different aspect dimensions based on their own preferences. Similarly, different items may contain different types of aspect information. It is therefore crucial to include user and item representations in the prompt construction process and make the aspect extraction process personalized.

\noindent\textbf{\emph{Constructing user and item representations.}} While there are many alternative methods to take into account feature information,  we follow the standard practice in deep neural network-based recommendation models  \cite{zhang2019deep} and map the users and items into latent embeddings. By doing so, we encode the preference information in the explicit user and item features through these latent embeddings.
These embeddings are then concatenated with the embeddings from the input review for personalized aspect extraction. 

\textbf{\emph{Fine-tuning Language Model.}} Before we utilize the LLMs for the aspect term extraction task, we first fine-tune its parameters to adapt the general language model to the specific use cases and datasets that we study in this paper. The fine-tuning process is conducted following the standard practice \cite{howard2018universal} by adding a fine-tuning layer to the top of the pre-trained LLM. The parameters of the pre-trained LLM will be fixed during the fine-tuning stage, as they contain high-level knowledge about the general probability distribution over sequences of words, while those of the fine-tuning layers will be optimized and updated to learn the domain-specific probability distribution over sequences of words in our user review datasets. The optimization is conducted by minimizing the perplexity of the next-word prediction task based on our review datasets, where the inputs consist of the first $n$ tokens of each review and the target output is the $n+1$-th token in the review. After the fine-tuning process, the parameters of the fine-tuning layer of the pre-trained GPT-2 model will be updated together with the soft prompt of user and item embeddings to perform our designated learning tasks (aspect term extraction and aspect-based recommendation). As we demonstrate through ablation studies, fine-tuning is an important step in our proposed model and contributes significantly to the performance of the aspect term extraction task. 

We now detail the design of the neural network to generate user and item embeddings as the soft prompt \cite{li2021prefix}. Specifically, the user and item IDs are mapped into the one-hot vector directly, and then fed into the embedding layer to obtain  the corresponding user embedding $W_{u} \in \mathcal{R}^{d_u}$ and item embedding $W_{i}\in \mathcal{R}^{d_i}$ accordingly. Meanwhile, we generate the review embedding $W_{R}$ by feeding the word sequence of original review text $R=[r_{1},r_{2}...]$ into the pre-trained language model. The soft prompt $P_{ui}$ is constructed by concatenating the user and item embeddings: $P_{ui}=[W_{u};W_i] \in \mathcal{R}^{d_u + d_i}$, which is further combined with the review embedding $W_R$ and fed into the pre-trained language model for personalized aspect extraction. As mentioned in the introduction, we also add additional fine-tuning layers to adjust the outputs of the language model to better fit the aspect term extraction tasks. The parameters of LLMs, along with these task-specific layers are fine-tuned on the target datasets before prompt tuning. The loss function of the prompt learning process in Component 1 is determined as a multi-label multi-class cross-entropy loss by comparing the predicted aspect terms $\hat{a}$ and the ground truth aspect terms $a$ as follows:
\begin{equation}
\label{eq:loss1}
    Loss_{extraction} = -\sum_{(u, i, R) \in \mathcal{D}}\sum_{k=1}^{K} I[\hat{a}=a_{k}] \times log[p(\hat{a}; u, i, R)]
\end{equation}
where $K$ is the number of ground truth aspect terms associated with the input tuple $(u, i, R)$ in the datasets and $p(\hat{a}; u, i, R)$ represents the probability that a certain aspect $\hat{a}$ is included in the given user review. Based on this loss function, we will be able to update the soft prompt and the associated user and item embedding tables through back-propagating, and the obtained aspects will be used for the subsequent recommendation process, which we will introduce in the next section.

\subsection{Component 2: Aspect-Based Recommendation}
\label{sec:aspect_based_recs}
After we obtain the most important personalized aspect terms in the previous stage, in this section we will perform aspect-based recommendations accordingly. Specifically, we adopt deep learning-based methods in providing aspect-based recommendations.

We now detail how we come up with the personalized aspect embedding, modulated by the user and item embeddings $W_u, W_i$. For each of the predicted aspect terms $\hat{a}$, we feed it into the embedding layer to construct the associate embeddings $W_{\hat{a}} \in \mathcal{R}^{d_a}$.
The attention network first concatenates the user and item embedding $[W_u; W_i]$ and feeds them into a feed-forward layer to produce a vector $Z_{u, i} \in \mathcal{R}^{d_a}$. The attention mask is then computed using a softmax function over the aspect embedding $W_{\hat{a}}$ and $Z_{u, i}$, following the attentive learning procedure described in \cite{guan2019attentive}:
\begin{equation}
attention_{u, i}(\hat{a})=softmax(W_{\hat{a}}*Z_{u, i})
\end{equation}
$attention_{u, i}(\hat{a}) \in \mathcal{R}^{d_a}$ is the soft attention value defined over the extracted aspect terms $\hat{a}$ for user $u$ and item $i$ in the recommendation process. The aspect embedding for aspect term $\hat{a}$ is then updated as 
\begin{equation}
W_{\hat{a}} = W_{\hat{a}} * attention_{u, i}(a)
\end{equation}
For each of the extracted aspect terms, we repeat the process. Finally, we concatenate all the user and item-modulated aspect embeddings together to form the aspect embedding $W_{u, i, \hat{A}} \in \mathcal{R}^{K\cdot d_a}$, where $K$ is the total number of extracted aspect terms.  This personalized aspect embedding  $W_{u, i, \hat{A}}$ is then concatenated with the user \& item feature embeddings $W_u, W_i$ that we obtain in the prompt learning phase to form the inputs to the aspect-based recommendation network to generate the final rating prediction, i.e., $\hat{y}_{u, i} = f([W_{u, i, \hat{A}}; W_u; W_i])$.
The entire aspect-based recommendation model is optimized by minimizing the following MSE (Mean Squared Error) loss function:
\begin{equation}
\label{eq:loss2}
    Loss_{recs} = \sum_{i=1}^{N} |y_{u, i}-\hat{y}_{u, i}|^{2}
\end{equation}
where $y_{u, i}$ is the ground truth rating of item $i$ given by user $u$, and $\hat{y}_{u, i}$ is the predicted rating generated by our aspect-based recommender system. We will then combine the two tasks together and perform the joint learning of components 1 \& 2, which we will introduce in the next section.
 
\subsection{Alternating Learning of Component 1 \& 2}
\label{sec:joint_learning}
As we have discussed in the introduction and related work section, it is important to combine the aspect term extraction and aspect-based recommendation tasks through joint training in order to produce aspect terms that are not only representative of the user reviews, but also useful for identifying the most relevant recommendations for the users. We will now describe how this is realized in the training procedure.

The alternating training procedure of our proposed model is described in Algorithm \ref{alg1} and works as follows. At the beginning of the training process for each offline dataset, we will first fine-tune the pre-trained large language model (in our case, GPT-2 \cite{radford2019language}) by learning and updating the parameters of the LLMs and the task-specific layers we added to minimize the cross-entropy loss for the next-word prediction tasks, following the standard practices of fine-tuning in most NLP applications \cite{howard2018universal}. The goal of fine-tuning is to make the language model more suitable for the given offline dataset and produce a better performance as a result. To avoid information leakage, the fine-tuning process is only conducted on the training data for each offline dataset. The parameters of the fine-tuned language model will then be fixed for the rest of the training process.

Then, we randomly initialize the user and item embedding tables following the standard normal distribution, and the aspect-based recommendation network following the uniform distribution between -0.1 and 0.1 at the beginning of the training. In the optimization process, we follow the alternating training technique \cite{takacs2012alternating,chavdarova2018sgan} to update the loss from the aspect-term extraction task and aspect-based recommendation task in an iterative manner until convergence. There are multiple benefits to adopting the alternating training procedure~\cite{chavdarova2018sgan}, such as improving model convergence and increasing the chance of jumping out of local maxima.
We demonstrate empirically through our three offline experiments that the proposed model can be trained efficiently through the alternating training approach and achieve good performance.

The alternating training works by first optimizing component 1 of the aspect term extraction task through the Loss Equation (\ref{eq:loss1}). In this step, the user and item embedding tables in the soft prompt will be updated, while the pre-trained language model, the fine-tuning layer, and the subsequent recommendation network will all remain fixed. Then in the second stage, we will optimize component 2 of the aspect-based recommendation task through the Loss Equation (\ref{eq:loss2}). In this step, we will update the recommendation network as well as user and item embedding tables. The pre-trained language model and the fine-tuning layer at this stage will remain fixed. We will then go back to the optimization of component 1, and so on and so forth. By alternating between these two objectives, our proposed model is expected to achieve these two tasks simultaneously: (1) generating personalized aspects, and (2) improving recommendation performance. The benefits of our proposed model will be empirically demonstrated through extensive offline experiments in the next section. 

\begin{algorithm}[ht]
\SetAlgoNoLine
\DontPrintSemicolon
\KwIn{Dataset $\mathcal{D} = \{(u, i, R, A)\}$, where $R$ is the review text and $A$ are the annotated aspect terms; Learnable embedding tables $e_u$, $e_i$ for all users $u=1,...,U$ and all items $i=1,...,I$;
Aspect-based recommendation network parameters $\theta_{recommendation}$; Learning rate $\alpha$; Training epochs $n_{epoch}$; Pre-trained Language Model $LM$} 
Fine-tune the pre-trained language model $LM$ on the targeted dataset and update its parameters. \;
Initialize the user and item embedding tables and the recommendation model. \;
Repeat the following until convergence: \;
 \For{$\forall (u, i, R, A) \in \mathcal{D}$}{
     Obtain review embeddings $W^R$ from user reviews $R$ based on the fine-tuned language model $LM$; \;
     Obtain user embeddings $W_{u}$ and item embeddings $W_{i}$ from user and item embedding tables $e_u$, $e_i$ respectively; \;
     Construct the soft prompt: $P_{ui} = [W_{u}, W_{i}]$; \;
     Generate the predicted list of aspects $\hat{A}_{ui}$ from the concatenated soft prompt $P_{ui}$ and review embedding $W_R$ using the fine-tuned language model $LM$; \;
     Update the embedding tables $e_u$, $e_i$ by back-propagating Loss Equation (\ref{eq:loss1}); \;
     Construct the recommendation inputs through concatenation: $Recommendation_{ui} = [W_{u}, W_{i}, \hat{A}_{ui}]$
     Generate the predicted ratings $\hat{r}_{ui}$ from the inputs $Recommendation_{ui}$ and the recommendation model, parameterized by network parameters $\theta_{recommendation}$; \;
     Updated network parameters $\theta_{recommendation}$ and the embedding tables by back-propagating Loss Equation (\ref{eq:loss2}).
     }
\KwOut{Generate the list of aspects $\hat{A}_{ui}$ and the predicted ratings $\hat{r}_{ui}$.}
\caption{Soft Prompt Tuning for Aspect-Based Recommendations}
\label{alg1}
\end{algorithm}

\section{Experiments}
\subsection{Data}
To demonstrate the benefits of our proposed model, we conduct extensive offline experiments on three large-scale industrial datasets \cite{li2022personalized} in three different recommendation applications, respectively from TripAdvisor (hotel), Amazon (movies \& TV), and Yelp (restaurant). Each record in these datasets is comprised of a user ID, an item ID, a rating on a scale of 1 to 5, a user review, and three aspect terms associated with that review as the ground truth. Some selected examples of user reviews and aspect terms are illustrated in Section 5.4. We have also listed the descriptive statistics of three datasets, as well as the sparsity levels (the total number of records divided by the total number of all user-item pairs) in Table \ref{statisticalnumber}. %\footnote{The model and code have been made publicly available at \url{https://anonymous.4open.science/r/Prompt-3708/}}

\begin{table}[h]
\centering
\begin{tabular}{|c|c|c|c|}
\hline
Dataset & \textbf{TripAdvisor} & \textbf{Amazon} & \textbf{Yelp}\\ \hline
\# of Ratings & 320,023 & 441,783 & 1,293,247 \\ \hline
\# of Users & 9,765 & 7,506 & 27,147 \\ \hline
\# of Items & 6,280 & 7,360 & 20,266 \\ \hline
Sparsity & 0.522\% & 0.800\% & 0.235\% \\ \hline
\end{tabular}
\newline
\caption{Descriptive Statistics of Three Datasets}
\label{statisticalnumber}
\end{table}

\subsection{Baselines, Metrics, and Experiment Settings}
To demonstrate the effectiveness of our proposed model, we compare its performance with selected state-of-the-art baselines for \textbf{aspect extraction} and \textbf{aspect-based recommendation}. The first group of baselines includes the following models for \textbf{aspect extraction}:

\begin{itemize}
\item \textbf{DE-CNN \cite{xu2018double}}, which is a novel and yet simple CNN-based model employing two types of pre-trained embeddings for the aspect term extraction task: general-purpose embeddings and domain-specific embeddings.
\item \textbf{LCFS \cite{phan2020modelling}}, which explores the grammatical aspect of the sentence and employs the self-attention mechanism for syntactical learning. It combines part-of-speech embeddings, dependency-based embeddings, and contextualized embeddings to enhance the performance of aspect term extraction.
\item \textbf{ABAE \cite{he2017unsupervised}}, which improves coherence by exploiting the distribution of word co-occurrences through the use of neural word embeddings. It also uses an attention mechanism to de-emphasize irrelevant words during training, further improving the coherence of aspects.
\item \textbf{BERT \cite{xu2020understanding}}, which leverages the annotated datasets in aspect term extraction tasks to investigate both the attentions and the learned representations of pre-trained BERT models.
\item \textbf{IMN \cite{he2019interactive}}, which is an interactive multi-task learning network that jointly learns multiple related tasks simultaneously at both the token and document levels. It introduces a message-passing architecture where information is iteratively passed to different tasks through a shared set of latent variables.
\item \textbf{JASA \cite{zhuang2020joint}}, which employs an autoencoder structure with the attention mechanism to learn two dictionary matrices for aspect and sentiment respectively. The aspect and sentiment encoders are jointly trained to enable sentiment embeddings in the dictionary to be tuned towards the aspect-specific sentiment words for each aspect, which benefits the classification performance.
\end{itemize}

The second group of baseline includes the following for \textbf{aspect-based recommendation}:
\begin{itemize}
\item \textbf{A3NCF \cite{cheng20183ncf}}, which is a new topic model to extract user preferences and item characteristics from review texts. It guides the representation learning of users and items, and also captures a user’s special attention on each aspect of the targeted item with an attention network.
\item \textbf{SULM \cite{bauman2017aspect}}, which first predicts the sentiment that the user may have about the item based on what he/she might express about the aspects of the item and then identifies the most valuable aspects of the user’s potential experience with that item. It further recommends items together with those most important aspects over which the user has control and can potentially select them.
\item \textbf{AARM \cite{guan2019attentive}}, which models the interactions between synonymous and similar aspects to enrich the aspect connections between user and product. It also contains a neural attention network to capture a user’s attention toward aspects when examining different products.
\item \textbf{MMALFM \cite{cheng2019mmalfm}}, which applies a multi-modal aspect-aware topic model to model users’ preferences and items’ features from different aspects, and also estimate the aspect importance of a user toward an item. The overall rating is then computed via a linear combination of the aspect ratings, which are weighted by the corresponding aspect importance.
\item \textbf{ANR \cite{chin2018anr}}, which performs aspect-based representation learning for both users and items via an attention-based component. It also models the multi-faceted process behind how users rate items by estimating the aspect-level user and item importance based on the neural co-attention mechanism.
\item \textbf{MTER \cite{le2021explainable}}, which provides comparative explanations involving such items, and also formulates comparative constraints involving aspect-level comparisons between the target item and the reference items.
\end{itemize}

To evaluate the performance of these two learning tasks in our experiments, we consider the following three metrics for determining the accuracy of aspect term extraction (by comparing with the ground truth aspect terms): Precision@3, Recall@3, F1-Score; and the following three metrics for understanding the effectiveness of the aspect-based recommendations: RMSE, MAE, AUC. For those aspect-based recommendation baselines that rely on explicit aspect information, we will use the ground truth aspects as the inputs. We then report the experimental results as the average of 10 independent runs.

To ensure a fair comparison between our proposed model and the baseline models, we identify the optimal hyperparameters for all selected models through Grid Search within the same amount of time (1 day). As a result, we formulate the fine-tuning layer of the pre-trained GPT-2 model as a fully-connected layer with an embedding size of 768. The recommendation network is formulated as three fully-connected layers with embedding size 128 and activation function of Sigmoid. The network parameters are optimized using the Stochastic Gradient Descent (SGD) technique.

\subsection{Ablation Models}
To further validate our model design and tease apart the importance of each component in our model, we also constructed a series of ablation models, and compared the performance of these variants with our proposed method. We summarize these ablation models below:
\begin{itemize}
\item \textbf{Ablation 1 (No Joint Training)}, where we train the tasks of aspect term extraction and aspect-based recommendation separately using the same loss function as in our proposed model.
\item \textbf{Ablation 2 (No Fine-Tuning)}, where we use the  pre-trained large language model directly to construct our proposed model, without first fine-tuning it on the three offline datasets.
\item \textbf{Ablation 3 (No Prompt)}, where we remove the soft prompt of user and item feature embeddings from the inputs, and only feed the review embeddings into the language model to generate the aspect terms.
\item \textbf{Ablation 4 (Discrete Prompt)}, where we replace the soft prompt of user and item ID embeddings by feeding them directly as discrete tokens, and along with the review text to the fine-tuned LLM.
\item \textbf{Ablation 5 (No Alternating Training)}, where we combine the loss of aspect term extraction and aspect-based recommendation together and update the model in one shot, without adopting the alternating training procedure.
\item \textbf{Ablation 6 (No Attention Mechanism)}, where we remove the attention mechanism from the aspect-based recommendation network.
\item \textbf{Ablation 7 (Only User Representation)}, where we use only the latent user representations as the soft prompt to feed into the language model for the aspect term extraction task.
\item \textbf{Ablation 8 (Only Item Representation)}, where we use only the latent item representations as the soft prompt to feed into the language model for the aspect term extraction task.
\end{itemize}

\section{Results}
\subsection{Main Results}
We present the results of our offline experiments in Table \ref{aspect_term_extraction} and Table \ref{aspect_recommendation}. As we can observe from these two tables, our proposed model significantly and consistently outperforms all the selected baseline models in terms of both the aspect term extraction and the aspect-based recommendation tasks. In particular, our proposed model could outperform the best baseline models by 2.57\% in Precision@3, 5.51\% in Recall@3, and 3.96\% in F1-Score for the aspect term extraction task, and by 4.08\% in RMSE, 4.89\% in MAE and 2.91\% in AUC for the aspect-based recommendation task in the Amazon dataset. Similar levels of performance improvements are also observed in the Yelp dataset and the TripAdvisor dataset.
On one hand, the significant increase of performance metrics over aspect term extraction baselines (Table \ref{aspect_term_extraction}) indicates that the aspect-based recommendation task is beneficial for the aspect term extraction task. Specifically, we would be able to learn more effective and useful aspect terms that are most indicative of user preference and experiences (i.e. more predictive of their ratings). On the other hand, the significant increase of performance metrics over aspect-based recommendation baselines (Table \ref{aspect_recommendation}) indicates that the extracted aspect terms are also helpful for the aspect-based recommendation task. The extracted aspect terms not only represent the essential information in user reviews, but are also helpful for recommendation tasks (rating prediction). Therefore, it is crucial to co-train the two tasks together as we proposed in this framework. Among the three datasets, the aspect extraction improvement on TripAdvisor is the smallest, although still significantly better than baselines, possibly due to the limited record number from this dataset. In addition, the soft prompt learning technique applied to the pre-trained large language model also contributes to the significant performance improvements, as we could observe from the results in Table \ref{aspect_term_extraction} and Table \ref{aspect_recommendation}, and the ablation studies from the next section.

To conclude, by utilizing the soft prompt learning technique and combining the training process for the two tasks, our proposed framework is able to generate more representative and useful aspects for recommendation purposes, leading to significant performance improvements in both tasks. We conduct additional experiments in the next section to further demonstrate the benefits of each component.

\begin{table*}
\centering
\begin{tabular}{|c|ccc|ccc|ccc|} \hline
Dataset & \multicolumn{3}{c|}{Amazon} & \multicolumn{3}{c|}{Yelp} & \multicolumn{3}{c|}{TripAdvisor} \\ \hline
Algorithm & Precision@3 & Recall@3 & F1-Score & Precision@3 & Recall@3 & F1-Score & Precision@3 & Recall@3 & F1-Score \\ \hline
\textbf{Our Model} & \textbf{0.2533*} & \textbf{0.2846*} & \textbf{0.2680*} & \textbf{0.2431*} & \textbf{0.2568*} & \textbf{0.2498*} & \textbf{0.2755*} & \textbf{0.2519*} & \textbf{0.2632*} \\ 
 & (0.0012) & (0.0011) & (0.0011) & (0.0011) & (0.0011) & (0.0011) & (0.0012) & (0.0011) & (0.0011) \\
(Improvement \%) & +2.57\% & +5.51\% & +3.96\% & +2.59\% & +2.73\% & +2.68\% & +0.98\% & +0.83\% & +0.91\% \\ \hline
DE-CNN & \underline{0.2468} & \underline{0.2689} & \underline{0.2574} & \underline{0.2368} & \underline{0.2498} & \underline{0.2431} & 0.2723 & 0.2496 & 0.2605 \\
 & (0.0019) & (0.0020) & (0.0017) & (0.0014) & (0.0013) & (0.0012) & (0.0017) & (0.0015) & (0.0012) \\ \hline
LCFS & 0.2449 & 0.2677 & 0.2558 & 0.2362 & 0.2496 & 0.2427 & 0.2705 & 0.2488 & 0.2592 \\
 & (0.0017) & (0.0016) & (0.0017) & (0.0014) & (0.0013) & (0.0013) & (0.0017) & (0.0017) & (0.0011) \\ \hline
ABAE & 0.2416 & 0.2650 & 0.2528 & 0.2350 & 0.2491 & 0.2418 & 0.2688 & 0.2471 & 0.2575 \\
 & (0.0023) & (0.0024) & (0.0026) & (0.0013) & (0.0013) & (0.0012) & (0.0016) & (0.0017) & (0.0012) \\ \hline
BERT & 0.2449 & 0.2681 & 0.2560 & 0.2359 & 0.2496 & 0.2426 & \underline{0.2728} & \underline{0.2498} & \underline{0.2608} \\
 & (0.0027) & (0.0024) & (0.0017) & (0.0013) & (0.0012) & (0.0011) & (0.0017) & (0.0017) & (0.0011) \\ \hline
IMN & 0.2430 & 0.2634 & 0.2528 & 0.2347 & 0.2481 & 0.2412 & 0.2715 & 0.2493 & 0.2599 \\
 & (0.0019) & (0.0021) & (0.0013) & (0.0013) & (0.0012) & (0.0011) & (0.0015) & (0.0012) & (0.0011) \\ \hline
JASA & 0.2408 & 0.2634 & 0.2516 & 0.2343 & 0.2481 & 0.2410 & 0.2691 & 0.2487 & 0.2585 \\ 
 & (0.0016) & (0.0022) & (0.0019) & (0.0014) & (0.0012) & (0.0011) & (0.0017) & (0.0011) & (0.0011) \\ \hline
Ablation 1 & 0.2420 & 0.2641 & 0.2526 & 0.2359 & 0.2498 & 0.2427 & 0.2688 & 0.2480 & 0.2580 \\
 & (0.0012) & (0.0011) & (0.0011) & (0.0013) & (0.0012) & (0.0011) & (0.0012) & (0.0011) & (0.0011) \\ \hline
Ablation 2 & 0.2485 & 0.2739 & 0.2606 & 0.2381 & 0.2515 & 0.2446 & 0.2726 & 0.2501 & 0.2609 \\
 & (0.0012) & (0.0011) & (0.0011) & (0.0011) & (0.0012) & (0.0011) & (0.0012) & (0.0011) & (0.0011) \\ \hline
Ablation 3 & 0.2428 & 0.2667 & 0.2542 & 0.2346 & 0.2498 & 0.2420 & 0.2680 & 0.2468 & 0.2570 \\
 & (0.0012) & (0.0011) & (0.0012) & (0.0011) & (0.0011) & (0.0011) & (0.0012) & (0.0011) & (0.0011) \\ \hline
Ablation 4 & 0.2428 & 0.2661 & 0.2539 & 0.2346 & 0.2491 & 0.2416 & 0.2685 & 0.2472 & 0.2574 \\
 & (0.0012) & (0.0012) & (0.0012) & (0.0013) & (0.0011) & (0.0011) & (0.0012) & (0.0011) & (0.0011) \\ \hline
Ablation 5 & 0.2496 & 0.2780 & 0.2631 & 0.2393 & 0.2538 & 0.2463 & 0.2736 & 0.2510 & 0.2618 \\
 & (0.0012) & (0.0012) & (0.0011) & (0.0013) & (0.0011) & (0.0011) & (0.0012) & (0.0011) & (0.0011) \\ \hline
Ablation 6 & 0.2498 & 0.2786 & 0.2634 & 0.2397 & 0.2541 & 0.2467 & 0.2738 & 0.2510 & 0.2619 \\ 
 & (0.0012) & (0.0012) & (0.0011) & (0.0011) & (0.0011) & (0.0011) & (0.0012) & (0.0011) & (0.0011) \\ \hline
Ablation 7 & 0.2493 & 0.2783 & 0.2627 & 0.2391 & 0.2537 & 0.2461 & 0.2730 & 0.2501 & 0.2615 \\ 
 & (0.0012) & (0.0012) & (0.0011) & (0.0011) & (0.0011) & (0.0011) & (0.0012) & (0.0011) & (0.0011) \\ \hline
Ablation 8 & 0.2488 & 0.2783 & 0.2631 & 0.2391 & 0.2538 & 0.2459 & 0.2726 & 0.2498 & 0.2613 \\ 
 & (0.0012) & (0.0012) & (0.0011) & (0.0011) & (0.0011) & (0.0011) & (0.0012) & (0.0011) & (0.0011) \\ \hline
\end{tabular}
\newline
\caption{Aspect term extraction performance in three datasets. `*' represents statistical significance with confidence level = 0.95. Improvement percentages are computed over the performance of the best baseline model for each metric.}
\label{aspect_term_extraction}
\end{table*}

\begin{table*}
\centering
\begin{tabular}{|c|ccc|ccc|ccc|} \hline
Dataset & \multicolumn{3}{c|}{Amazon} & \multicolumn{3}{c|}{Yelp} & \multicolumn{3}{c|}{TripAdvisor} \\ \hline
Algorithm & RMSE & MAE & AUC & RMSE & MAE & AUC & RMSE & MAE & AUC \\ \hline
\textbf{Our Model} & \textbf{0.2083*} & \textbf{0.1757*} & \textbf{0.7243*} & \textbf{0.2413*} & \textbf{0.2053*} & \textbf{0.6991*} & \textbf{0.1975*} & \textbf{0.1709*} & \textbf{0.7071*} \\ 
 & (0.0011) & (0.0009) & (0.0017) & (0.0011) & (0.0009) & (0.0016) & (0.0011) & (0.0009) & (0.0017) \\
(Improvement \%) & +4.08\% & +4.89\% & +2.91\% & +6.80\% & +4.43\% & +2.59\% & +5.62\% & +5.38\% & +2.39\% \\ \hline
A3NCF & 0.2246 & 0.1895 & 0.6964 & 0.2611 & 0.2176 & 0.6780 & 0.2108 & 0.1814 & 0.6875 \\
 & (0.0013) & (0.0010) & (0.0022) & (0.0014) & (0.0011) & (0.0018) & (0.0012) & (0.0009) & (0.0021) \\ \hline
SULM & 0.2478 & 0.1977 & 0.6851 & 0.2825 & 0.2255 & 0.6612 & 0.2199 & 0.1874 & 0.6733 \\
 & (0.0013) & (0.0010) & (0.0024) & (0.0013) & (0.0010) & (0.0018) & (0.0012) & (0.0010) & (0.0023) \\ \hline
AARM & \underline{0.2168} & \underline{0.1843} & \underline{0.7032} & 0.2589 & 0.2159 & 0.6799 & 0.2089 & 0.1805 & 0.6898 \\
 & (0.0012) & (0.0009) & (0.0024) & (0.0014) & (0.0010) & (0.0018) & (0.0011) & (0.0009) & (0.0023) \\ \hline
MMALFM & 0.2305 & 0.1930 & 0.6928 & 0.2596 & 0.2163 & 0.6801 & 0.2120 & 0.1822 & 0.6892 \\
 & (0.0012) & (0.0009) & (0.0019) & (0.0013) & (0.0010) & (0.0016) & (0.0011) & (0.0009) & (0.0023) \\ \hline
ANR & 0.2277 & 0.1915 & 0.6958 & \underline{0.2577} & \underline{0.2144} & \underline{0.6810} & \underline{0.2086} & \underline{0.1801} & \underline{0.6902} \\
 & (0.0012) & (0.0009) & (0.0017) & (0.0013) & (0.0010) & (0.0016) & (0.0011) & (0.0009) & (0.0021) \\ \hline
MTER & 0.2286 & 0.1903 & 0.6964 & 0.2621 & 0.2163 & 0.6801 & 0.2101 & 0.1827 & 0.6885 \\ 
 & (0.0011) & (0.0009) & (0.0019) & (0.0013) & (0.0009) & (0.0016) & (0.0011) & (0.0009) & (0.0021) \\ \hline
Ablation 1 & 0.2250 & 0.1900 & 0.6980 & 0.2568 & 0.2141 & 0.6825 & 0.2081 & 0.1801 & 0.6933 \\
 & (0.0012) & (0.0009) & (0.0017) & (0.0011) & (0.0009) & (0.0016) & (0.0011) & (0.0009) & (0.0017) \\ \hline
Ablation 2 & 0.2142 & 0.1799 & 0.7197 & 0.2440 & 0.2090 & 0.6962 & 0.2001 & 0.1741 & 0.7045 \\
 & (0.0011) & (0.0009) & (0.0017) & (0.0011) & (0.0009) & (0.0016) & (0.0011) & (0.0009) & (0.0017) \\ \hline
Ablation 3 & 0.2398 & 0.1942 & 0.6903 & 0.2677 & 0.2189 & 0.6784 & 0.2144 & 0.1886 & 0.6855 \\
 & (0.0012) & (0.0009) & (0.0016) & (0.0011) & (0.0009) & (0.0016) & (0.0011) & (0.0009) & (0.0017) \\ \hline
Ablation 4 & 0,2375 & 0.1926 & 0.6915 & 0.2661 & 0.2180 & 0.6776 & 0.2140 & 0.1867 & 0.6877 \\
 & (0.0011) & (0.0009) & (0.0017) & (0.0011) & (0.0009) & (0.0016) & (0.0011) & (0.0009) & (0.0017) \\ \hline
Ablation 5 & 0.2298 & 0.1917 & 0.6966 & 0.2581 & 0.2152 & 0.6801 & 0.2095 & 0.1844 & 0.6898 \\
 & (0.0011) & (0.0009) & (0.0017) & (0.0011) & (0.0009) & (0.0016) & (0.0011) & (0.0009) & (0.0017) \\ \hline
Ablation 6 & 0.2196 & 0.1820 & 0.7158 & 0.2479 & 0.2117 & 0.6844 & 0.2059 & 0.1770 & 0.6967 \\ 
 & (0.0011) & (0.0009) & (0.0017) & (0.0011) & (0.0009) & (0.0016) & (0.0011) & (0.0009) & (0.0017) \\ \hline
Ablation 7 & 0.2173 & 0.1811 & 0.7179 & 0.2463 & 0.2111 & 0.6856 & 0.2052 & 0.1768 & 0.6971 \\ 
 & (0.0011) & (0.0009) & (0.0017) & (0.0011) & (0.0009) & (0.0016) & (0.0011) & (0.0009) & (0.0017) \\ \hline
Ablation 8 & 0.2178 & 0.1812 & 0.7172 & 0.2468 & 0.2110 & 0.6862 & 0.2052 & 0.1768 & 0.6970 \\ 
 & (0.0011) & (0.0009) & (0.0017) & (0.0011) & (0.0009) & (0.0016) & (0.0011) & (0.0009) & (0.0017) \\ \hline
\end{tabular}
\newline
\caption{Aspect-based recommendation performance in three datasets. `*' represents statistical significance with confidence level = 0.95. Improvement percentages are computed over the performance of the best baseline model for each metric.}
\label{aspect_recommendation}
\end{table*}

\subsection{Ablation Study}
%\minminc{pls make one paragrah for each ablation study, and discuss in more details. }
Apart from the comparison with state-of-the-art baseline models, we also conducted extensive ablation studies to compare with different variants of our proposed model to demonstrate the value of each component in our proposed framework. The results are listed in Table \ref{aspect_term_extraction} and Table \ref{aspect_recommendation} respectively. By comparing the performance of our proposed model and the ablation models, we have the following observations: 

\paragraph{Ablation 1: No joint training:} In the Ablation 1 model, we demonstrate that the tasks of aspect term extraction and aspect-based recommendations are mutually beneficial for each other, as training them separately would result in a significant performance decrease (around 5\%) in both tasks, as shown in Table \ref{aspect_term_extraction} and \ref{aspect_recommendation}.
%\minminc{can you find some examples showing the aspects extracted in joint training vs separate?}
As an example shown in Table \ref{casestudy1}, our proposed joint training model extracts the most important aspect terms that match perfectly with the ground truth, while the separate training model could only extract the aspect terms as "Family, Movie, Good". The term "freakish" is especially important for the recommendation process and it connects the targeted movie with other types of movies, such as comics or thrillers.

\begin{table}[h]
\centering
\resizebox{0.75\textwidth}{!}{
\begin{tabular}{|c|c|} \hline
Original & "It was wonderful to have a family movie with very few expletives despite the \\
Review & (comically) freakish nature of the Adams family. Overall this is a very good family movie." \\ \hline
Ground Truth & Family, Movie, Freakish \\ \hline
Our Model & Family, Movie, Freakish \\ \hline
Ablation 1 & Family, Movie, Good \\ \hline
\end{tabular}
}
\newline
\caption{Showcase of the aspect term extraction task for Ablation 1}
\label{casestudy1}
\end{table}

\paragraph{Ablation 2: No fine-tuning:} In the Ablation 2 model, we directly use the pre-trained GPT-2 model to perform the aspect term extraction task, and the parameters of the GPT-2 model remain fixed throughout the entire training process. We observed a significant performance loss (over 2 \%) for both the aspect term extraction and the aspect-based recommendation task. we conclude that fine-tuning is an important step in our proposed model. With fine-tuning, the large language model is adapted to better suit the specific use case and the data associated with it, which explains its benefits in improving the aspect extraction task (Table \ref{aspect_term_extraction}). This observation corroborates with the common practice to use fine-tuning to align LLMs to downstream tasks. As the aspect extraction task and aspect-based recommendation task are jointly trained, this explains the improvement in the aspect-based recommendation task as well (Table \ref{aspect_recommendation}). 

\paragraph{Ablation 3: No prompt:} In the Ablation 3 model, we do not include the continuous prompt component and feed only the review embeddings as input to the LLM to extract aspect terms. In that sense, the LLM would not be able to learn the relevant user information (such as writing style, the previous purchasing experience, etc..) and item information (such as product content, category, etc..) that both play important roles in the aspect term extraction task. As a result, this ablation model performs significantly worse (around 5\%) than our proposed model in both the aspect extraction task and aspect-based recommendation task, and across all three datasets, showing the importance of personalization. 
%\minminc{Again show examples.}
As an example shown in Table \ref{casestudy3}, our proposed model could extract the important aspect term "Sweetness" from the review, while without the prompt inputs, the ablation model would only extract the aspect terms of "Cleanliness, Environment, Coffee". The former is preferable in this case, as we could observe based on the consumer's previous reviews that she/he has valued the taste of sweetness a lot when making decisions for places to hang out.
%\yuyanw{Instead of verbalizing them, let's create a table (similar to Table 4) for these examples from the alation studies?}

\begin{table}[h]
\centering
\resizebox{0.75\textwidth}{!}{
\begin{tabular}{|c|c|} \hline
Original & "I have been to a few great places and honestly they all taste similar but the cleanliness and environment are unbeatable. \\
Review & Also, the coffee was prevalent and the chocolate was enough to add sweetness without making it cloyingly sweet." \\ \hline
Ground Truth & Cleanliness, Environment, Sweetness \\ \hline
Our Model & Cleanliness, Environment, Sweetness \\ \hline
Ablation 3 & Cleanliness, Environment, Coffee \\ \hline
\end{tabular}
}
\newline
\caption{Showcase of the aspect term extraction task for Ablation 3}
\label{casestudy3}
\end{table}

\paragraph{Ablation 4: Discrete Prompt:} In the Ablation 4 model, we replace the continuous prompt of user and item embeddings with the discrete prompt. In other words, every user and item is encoded as a discrete ID token which is fed into the LLM as input. The LLM will be able to update its understanding of the ID tokens during the training and backpropagation process, as these tokens should provide additional information on the aspect distribution within each review. We also observe significant performance loss (around 5\%) for ablation model 4. This confirms the effectiveness of user/item embeddings as well as the continuous/soft prompt technique in our proposed model. In addition, it is important and necessary to make these prompts trainable to achieve better performance for both tasks.
%\minminc{talk about the how LLMs will be able to recognize these "new" discrete tokens.}

\paragraph{Ablation 5: No Alternating Training:} In the Ablation 5 model, we do a universal back-propagation to all the trainable parameters for both aspect extraction and aspect-based recommendation task by adding the two losses together. We observe a significant performance loss (around 2\%) due to this lack of THE alternating training mechanism. This demonstrates the benefits and necessity of the alternating training mechanism, which enables the end-to-end framework to first improve its performance on one learning task, and then the other in an alternating and iterative manner, until both tasks converge and achieve the optimal performance. 

\paragraph{Ablation 6: No Attention Mechanism:} In the Ablation 6 model, we remove the attention used for the aspect-based recommendation framework. There is a small but significant drop in the performance of both the aspect extraction and the aspect-based recommendation task across all three datasets (Table \ref{aspect_term_extraction} and \ref{aspect_recommendation}). This validates the importance of the attention mechanism in our proposed model.
%\minminc{is it possible to add some visualization on the attention weights? maybe showing different users/items attends to different things.} 
%\pan{I might not have enough time to re-run the experiments and record the attention values.}

\paragraph{Ablation 7 \& 8: No item representation \& no user representation:} In the Ablation 7 and 8 models, we remove either latent user representations or latent item representations from the soft prompt of the language model, as opposed to ablation 3 where both are removed. We again witness a similar and significant performance decrease (around 2\%) for both ablation models, in both the aspect term extraction and aspect-based recommendation tasks(Table \ref{aspect_term_extraction} and \ref{aspect_recommendation}). Ablation 7 shows the benefit of personalization in the soft prompts for aspect extraction. Ablation 8 shows the benefit of item-specific aspects. For example, different items reviewed by the same user might differ in the characteristics of the extracted aspects. 
%\minminc{talk about how the results compare to ablation 3. if removing both has bigger effect than the combination of both, explain why.}
Another interesting observation is that adding only item representation or only user representation as the prompt to the language model actually achieves inferior performance than the one in Ablation 3, where we remove both user and item representations from the prompt. One possible explanation is that both aspect term extraction and aspect-based recommendation tasks require the \textbf{mutual} understanding of user preferences and item content and in particular their interactions, while adding partial information to the model might involve more noise than useful signals, therefore backfiring in terms of model performance.

To sum up, as there are multiple moving parts in our proposed end-to-end framework, we conducted an extensive set of ablation studies to validate the importance of each component. All these additional results suggest that each component in our proposed model is useful and contributes to significant performance improvements in both aspect term extraction and aspect-based recommendation applications.

\subsection{Scalability Analysis}
\begin{figure}
\centering
\includegraphics[width=0.5\textwidth]{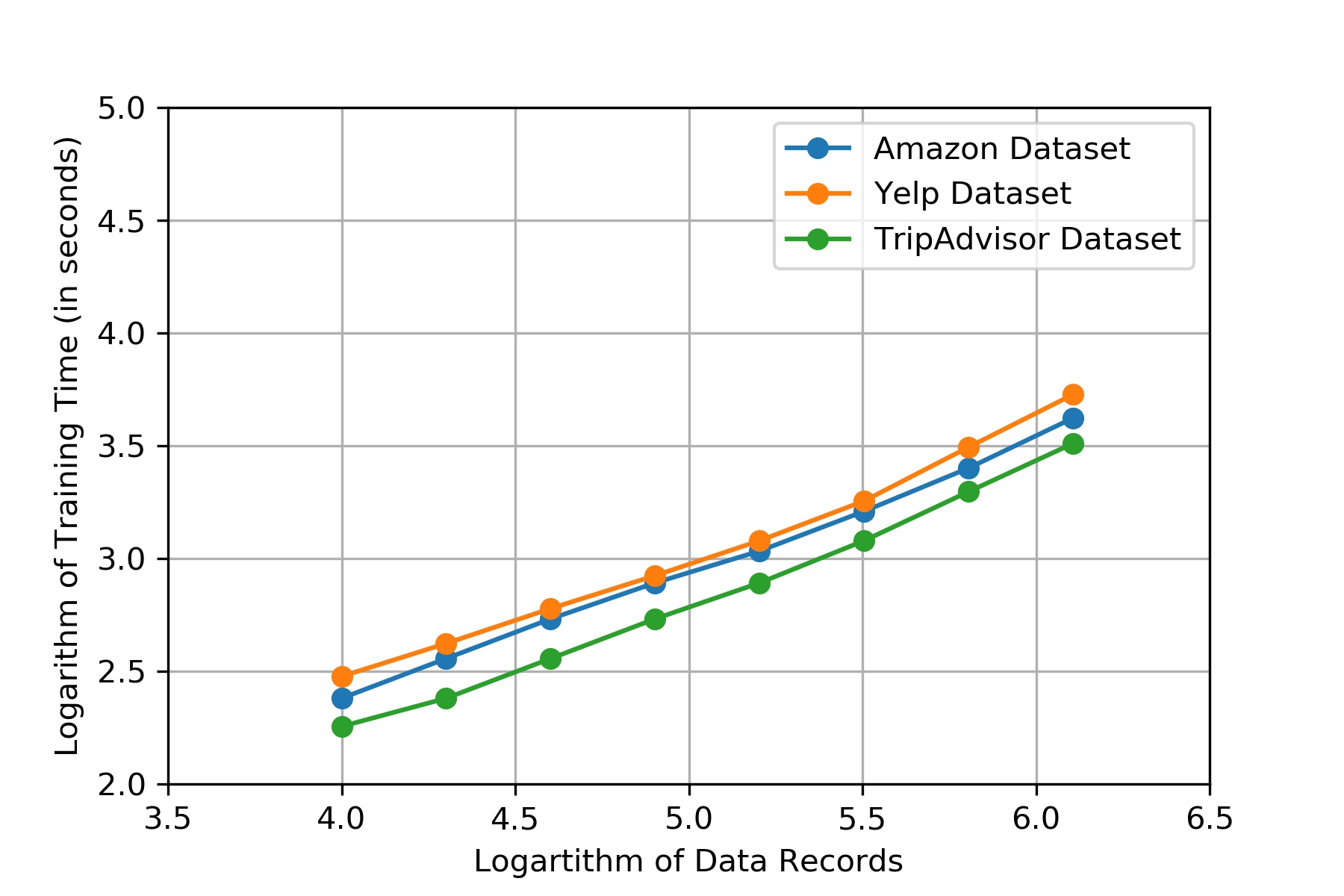}
\caption{Scalability Analysis in three datasets using the training time for completing 10 epochs}
\label{scalability}
\end{figure}

To further demonstrate the practicability of our proposed method, we also conduct a set of scalability analyses in this section, where we train our method on a series of randomly sampled subsets of the three offline datasets with sizes ranging from 10,000 to 1,280,000 in the number of records. The training is conducted on an MX450 GPU, and we compare the required training time for different subsets respectively. As we could observe from Figure \ref{scalability}, the training time of our method scales linearly with the number of records in the training data. In addition, we also verify that there are no statistical differences between the required training time of our model and the baseline models. Since we fix the parameters of the pre-trained and fine-tuned LLMs and only prompt tuning the model, the computational cost is relatively low. At the same time, our model could achieve significantly better performance in both aspect term extraction and aspect-based recommendation tasks, further illustrating the practical advantages of our proposed model. Finally, we would like to point out that our model could be scaled to more users and items through clustering certain users and items in the feature embedding generation process, following the techniques discussed in \cite{kang2020learning}.

\begin{table}[h]
\centering
\resizebox{0.5\textwidth}{!}{
\begin{tabular}{|c|c|} \hline
Original & "It is a great collection version of star wars original episodes \\
Review 1 & and worth purchasing through amazon if you are a fan." \\ \hline
Ground Truth & Star Wars, Original, Worth \\ \hline
Our Model & Star Wars, Original, Worth Purchasing \\
DE-CNN & Collection, Star Wars, Episode \\
LCFS & Star Wars, Worth, Amazon \\
ABAE & Collection, Episode, Worth \\ \hline
 & \\ \hline
Original & "This movie is still a wonderful adventure \\
Review 2 & which stands up well to the test of time." \\ \hline
Ground Truth & Wonderful, Adventure, Test of Time \\ \hline
Our Model & Wonderful, Adventure, Test of Time \\
DE-CNN & Movie, Wonderful, Well \\
LCFS & Movie, Wonderful, Adventure \\
ABAE & Movie, Wonderful, Time \\ \hline
 & \\ \hline
Original & "The bathroom looked a little dated \\
Review 3 & and the water pressure was on the low end." \\ \hline
Ground Truth & Bathroom, Dated, Low End \\ \hline
Our Model & Bathroom, Dated, Water Pressure \\
DE-CNN & Bathroom, Dated, Low \\
LCFS & Bathroom, Little, Water \\
ABAE & Bathroom, Little, Dated \\ \hline
\end{tabular}
}
\newline
\caption{Case study of the aspect term extraction task}
\label{casestudy}
\end{table}

\subsection{Case Study}
Finally, we present a set of case studies to showcase the benefits of our proposed model. Table \ref{casestudy} presents three example reviews from the test set and their ground truth aspect annotations, as well as aspects extracted from several baselines and our proposed method. 
One can see that our proposed model could effectively select the most important aspect terms from the original reviews, which are not only closest to the ground truth but also more useful for the subsequent recommendation tasks, as they represent the essence of user preference information. For example, in the first review, our model could identify the topic of the product on "Star Wars“, its most important characteristic "Original" and the user feedback on the product "Worth Purchasing". Other baseline models, however, cannot extract all three important aspects at the same time. We can also observe from the other two examples that our method performs significantly better than baseline models in aspect term extraction accuracy, which again demonstrates the advantages of joint training the aspect term extraction and aspect-based recommendation tasks in our proposed model.

\section{Conclusions and Future Work}
Providing aspect-based explanations and recommendations to the users is an important task in the design of recommender systems. While existing methods focus either on aspect term extraction or aspect-based recommendation tasks, we argue in this paper that it is important and beneficial to connect the learning process of these two tasks (i.e. aspect term extraction and aspect-based recommendation) in an end-to-end manner. Specifically, we propose an end-to-end aspect-extraction and aspect-based recommendation framework that extracts personalized aspect terms that are not only representative of the user reviews, but also useful for identifying the most relevant recommendations for the users. In addition, we leverage pre-trained large-language models to improve the quality of the extracted aspects.

The proposed framework adopts a personalized prompt-tuning mechanism, where the soft prompt is constructed based on user and item feature embeddings, which are concatenated with the review embeddings as the inputs to the pre-trained and fine-tuned language model. The output of the language model is a list of aspect terms, which are fed into the attentive neural network to generate aspect-based recommendations. Extensive offline experiments on three real-world datasets demonstrate the benefits of our proposed model, where it significantly outperforms several state-of-the-art baseline methods in both aspect-term extraction and aspect-based recommendation tasks.

Our method currently relies on user and item IDs to encode user and item preferences in aspect extraction and aspect-based recommendation. It however can be easily extended to use user and item features to address the cold-start problem and scale to a large number of users and items. In future work, we plan to further study the embedding generation process in the soft prompt design to make them more suitable for the aspect term extraction and the subsequent recommendation tasks in industrial recommendation settings by utilizing user and item features. We also plan to work on better fine-tuning strategies to make the large-scale pre-trained language model more useful for the designated learning applications.

\bibliographystyle{ACM-Reference-Format}
\bibliography{sample-base}

\end{document}